\documentclass[journal,twoside,web, 10pt]{ieeecolor}
\usepackage{generic}
\usepackage{cite}
\usepackage{amsmath,amssymb,amsfonts}
\usepackage{algorithmic}
\usepackage{graphicx}
\usepackage{multirow}
\usepackage{ragged2e}
\usepackage{graphicx}
\usepackage{textcomp}
\usepackage{makecell}
\usepackage{booktabs}
\usepackage{amsmath}
\usepackage{siunitx}
\usepackage{array}
\usepackage{caption}
\usepackage{subcaption}
\usepackage{stfloats}
\usepackage{url}

\makeatletter
\AtBeginDocument{%
  \def\doi#1{\url{https://doi.org/#1}}}
\makeatother

\makeatletter
\providecommand\add@text{}
\newcommand\tagaddtext[1]{%
  \gdef\add@text{#1\gdef\add@text{}}}%
\renewcommand\tagform@[1]{%
  \maketag@@@{\llap{\add@text\quad}(\ignorespaces#1\unskip\@@italiccorr)}%
}
\makeatother

\newcolumntype{M}[1]{>{\centering\arraybackslash}m{#1}}
\newcolumntype{E}[1]{>{\centering\arraybackslash\columncolor{Col}}m{#1}}
\newcolumntype{R}[1]{>{\raggedleft\arraybackslash}p{#1}}
\newcolumntype{L}[1]{>{\raggedright\arraybackslash}p{#1}}

\def\BibTeX{{\rm B\kern-.05em{\sc i\kern-.025em b}\kern-.08em
    T\kern-.1667em\lower.7ex\hbox{E}\kern-.125emX}}
\begin{document}
\bstctlcite{IEEEexample:BSTcontrol}

\title{Sub-Scalp Brain-Computer Interface Device Design and Fabrication}
\author{Timothy B. Mahoney, David B. Grayden, and Sam E. John
\thanks{T.B. Mahoney, D.B. Grayden, and S.E. John are with Dept of Biomedical Engineering, University of Melbourne, Victoria 3010, Australia. ({\tt tbmahoney@student.unimelb.edu.au})}%
\thanks{T.B. Mahoney is supported by the Elizabeth and Vernon Puzey Scholarship.}%
\thanks{D.B. Grayden is with Graeme Clark Institute, University of Melbourne, Victoria 3010, Australia.}}%

\maketitle

\begin{abstract}
Current brain-computer interfaces (BCI) face limitations in signal acquisition. While sub-scalp EEG offers a potential solution, existing devices prioritize chronic seizure monitoring and lack features suited for BCI applications. This work addresses this gap by outlining key specifications for sub-scalp BCI devices, focusing on channel count, sampling rate, power efficiency, and form factor. We present the Set-And-Forget EEG (SAFE) system, a custom-built amplifier and wireless transmitter meeting these criteria. This compact (12$\times$12~mm), six-channel device offers 1024~Hz sampling and Bluetooth Low Energy data transmission. Validation using generated sinusoids and electrocorticography recordings of visual evoked potentials in sheep models demonstrated low noise recording. Future animal studies will assess sub-scalp EEG signal quality for BCI applications. This data lays the groundwork for human trials, ultimately paving the way for chronic, in-home BCIs that empower individuals with physical disabilities.
\end{abstract}


\section{Introduction}
\label{sec:introduction}
\IEEEPARstart{B}{}rain-computer interfaces (BCIs) are systems that aim to establish a direct communication link between neural activity in the brain and an external device. Potential BCI applications span the fields of rehabilitation \cite{riccio_chapter_2016}, communication \cite{birbaumer_spelling_1999}, business \cite{arico_chapter_2016}, and entertainment \cite{beveridge_chapter_2016}. However, persons living with debilitating ailments causing severe paralysis stand to benefit the most from this technology \cite{chaudhary_chapter_2016,khan_review_2020, kubler_user-centered_2014, mcfarland_brain-computer_2020}. Conditions such as motor neurone disease, with a global incidence of 64,000 people per year \cite{logroscino_global_2018}, can leave patients with severe paralysis. For these people, BCI systems offer improved quality of life by facilitating communication and interaction with the world around them.

Current BCI technology has limitations preventing long-term, in-home use. Invasive implants, such as electrocorticography (ECoG) and penetrating arrays, are accompanied by significant risks to the patient during both implantation and chronic use, and can result in scaring that damages neural tissue and attenuates signal quality over time \cite{johnston_complications_2006, nagahama_intracranial_2019, onal_complications_2003, rolston_national_2015, taussig_invasive_2012, winslow_comparison_2010, winslow_quantitative_2010, theunisse_risk_2018}. Non-invasive electroencephalography (EEG) involves lengthy donning and doffing procedures, suffers from poor signal stability and quality, and are regarded as unaesthetic and cumbersome by potential BCI users \cite{kubler_user-centered_2014, miralles_braincomputer_2015, rashid_current_2020}. Endovascular (EV) stent-electrode arrays have recently attempted to address these limitations \cite{oxley_motor_2021}; however, these devices have poor spatial coverage (being confined to brain regions with large endovascular structures), cannot be removed once implanted, and may lead to adverse effects seen with similar stenting procedures \cite{starke_endovascular_2015}. There is a need for a minimally invasive signal acquisition method that addresses the disadvantages with current technologies and that will enable widespread long-term BCI use.

Placement of electrodes into the sub-scalp (also termed sub-galeal or sub-dermal) space offers a viable solution for BCI that overcomes many of the disadvantages of current signal acquisition methods.  Unlike EEG caps, sub-scalp devices can be discrete and wireless and require no donning and doffing procedures. They do not require risky brain surgery to be implanted. Sub-scalp electrodes can be positioned as needed across the skull, providing access to activity from brain regions outside the reach of EV arrays and can be easily removed if needed for replacement or upgrades.  The safety of chronic implantation of electronics beneath the scalp has been demonstrated by cochlear implants around the world over decades \cite{jiang_analysis_2017, weder_management_2020, desasouza_cochlear_2022}. Sub-scalp EEG is currently being employed for seizure monitoring and has shown promise for long-term use \cite{duun-henriksen_new_2020, stirling_seizure_2021, weisdorf_ultra-long-term_2019}. Current evidence of safety and stability provide hope for sub-scalp BCIs to record high quality neural signal for decoding and control of external devices. However, these devices typically only include a small number of electrodes ($\le$8) and low sampling rates ($\le$250~Hz), that are inappropriate for BCI applications. In this study, we have formulated a set of optimal specifications for sub-scalp BCI devices, fabricated a device to meet these specifications, and validated the device's performance compared to a state of the art EEG system.

High number of channels and sampling frequency are important for BCI applications. Paradigms for BCI control most often either rely on the subject attending a visual stimulus (such as P300 or steady-state visual evoked potential, i.e. SSVEP), or modulating cortex activity associated with motor function (such as motor imagery). Electrodes must be positioned appropriately over brain regions to record these paradigms. An implanted BCI system that includes electrodes positioned to record both of these phenomena would allow for a versatile interface, that could be operated via motor imagery, visual stimuli, or both. Visual stimulus BCIs have been demonstrated with two-channel EEG systems, with electrodes situated about the occipital lobe \cite{autthasan_single-channel_2020, cecotti_robust_2011, colwell_channel_2014, juan_alberto_brain-computer_2021, karunasena_single-channel_2021, mccann_electrode_2015, speier_method_2015, xu_channel_2013}. Motor BCIs can function with as few as two EEG channels over the motor cortex \cite{lun_simplified_2020}, although 8-10 electrodes surrounding C3 and C4 (international 10-20 system) is more commonly considered optimal \cite{arvaneh_optimizing_2011, lal_support_2004, tam_minimal_2011}. 

Regarding sampling frequency, high sampling allows for recording of high frequency EEG. High frequency EEG signals have demonstrated utility for BCI applications \cite{grosse-wentrup_braincomputer_2014, guger_utilizing_2013, loza_classification_2014}, and sub-scalp EEG may be capable of detecting high frequency brain oscillations up to 200~Hz \cite{olson_comparison_2016}. As such, the analogue to digital sampling frequency of the sub-scalp BCI must be high enough to capture such signals to maximise BCI functionality. 

In addition to electrode count and sampling frequency, there are several other design considerations for a sub-scalp BCI system. These considerations can compete, presenting trade offs. Since the system is implanted, data and power must be transmitted wirelessly. High sampling rate over numerous channels results in high data transmission throughput requirements and high power consumption. However, as we are developing a chronic, battery-operated implant, minimisation of power consumption can lead to a longer lasting battery. Additionally, the implanted component must be small and lightweight to minimise risk and discomfort to the user. These characteristics (electrode count, sampling frequency, power consumption, data transmission, size, and weight) must be balanced appropriately to realise a functional sub-scalp BCI. 

Here, we clearly define design characteristics of sub-scalp BCI devices. We then present the Set-And-Forget EEG (SAFE) system, a sub-scalp device designed to meet these specifications. A prototype was built, validated against the specifications, and compared with a commercially available state-of-the-art electrophysiology amplifier system in a visual evoked potential (VEP) study in sheep. The device includes commercially available components, and utilises Bluetooth Low Energy (BLE) for reliable and low-power electrophysiology data transmission from within the body to a computer for decoding, while occupying a custom printed circuit board (PCB) small enough to fit within current cochlear implant hermetic cans \cite{uwe_baumann_device_2020, patrick_development_2006}. Rather than rely on state-of-the-art laboratory EEG recording systems, we will instead use this device in animal experiments to investigate sub-scalp EEG signal quality, providing closer resemblance to the signal quality of chronic, in-home sub-scalp BCIs of the future. Results from these experiments will provide support for long-term, in-human sub-scalp BCI clinical studies that hopefully propel this technology toward regulatory approval and availability for persons with severe paralysis, providing them with a means of interacting with their environment. 

\section{Methods}
\subsection{Sub-scalp EEG (SAFE) Device Design and Fabrication}
The SAFE system was designed to fit certain specifications in the interests of safety and efficacy, and fabricated to meet these specifications with commercially available components. The key constraints guiding design decisions were minimisation of power consumption to prolong operating time, minimisation of package size to allow for implantation, and sufficient data quality for BCI applications in terms of channel count, resolution, sampling frequency, and transmission frequency. The full list of specifications, justification, and the characteristics of the SAFE system are discussed in Section~\ref{sec:CH2_Results}.

\subsection{Validation Procedures}
The SAFE system was validated by comparison with the state-of-the-art electrophysiology amplifier unit g.USBAmp (g.tec medical engineering, Austria). The amplifiers were compared in two experiments: \textit{in vitro} and \textit{in vivo}. First, a function generator provided sinusoidal voltage inputs to both systems simultaneously, and the error was compared between the recordings and a true sinusoid. Then, we provided a visual stimulus to sheep models, recorded ECoG visual evoked potentials (VEPs), and compared signal qualities between the two systems. All analyses were performed in MATLAB (Version 2023a, MathWorks, USA). All recordings by the g.USBAmp amplifier used a sampling frequency of 1200~Hz per channel, whereas the SAFE system sampled at 1024~Hz per channel.

\subsubsection{Sinusoidal Recordings}
\begin{figure}[tb]
    \includegraphics[width=0.48\textwidth]{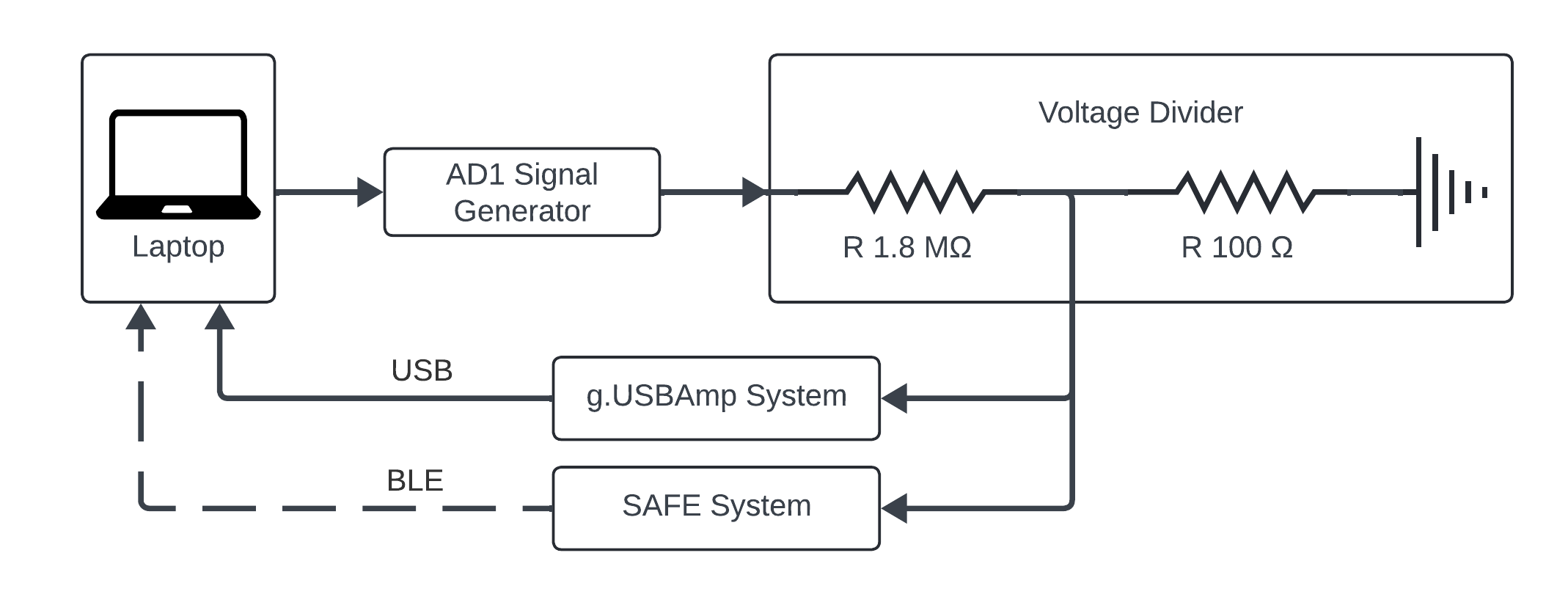}
    \centering
    \caption[Sinusoidal Recording Experiment Diagram]{An Analog Discovery 1 supplies sinusoidal voltages to the voltage divider circuit. The g.USBAmp and SAFE systems simultaneously sample the divided voltage and transmit the data to the PC via USB (g.USBAmp) and BLE (SAFE system).}
    \label{fig:Fig1}
\end{figure}

Sinusoidal waves with frequencies ranging over 10-190~Hz (10~Hz intervals with constant 50~µV amplitude) and amplitudes ranging over 10-100~µV (10~µV intervals with constant 20~Hz frequency) were generated, one at a time, using an Analog Discovery 1 (Digilent, USA) and were recorded by both amplifiers simultaneously for a 30 s period (Fig.~\ref{fig:Fig1}). A voltage divider was used to step down the signal to the microvolt range. These frequencies and amplitudes were chosen as they reflect the ranges of interest for sub-scalp EEG recordings. A voltage divider circuit was used to step down the signal from the Analog Discovery device's output range to the required amplitude range. The resistors used in the divider were the same for each recording and were assumed to be exactly their labeled resistance.

\begin{figure}
    \centering
    \includegraphics[width=0.48\textwidth]{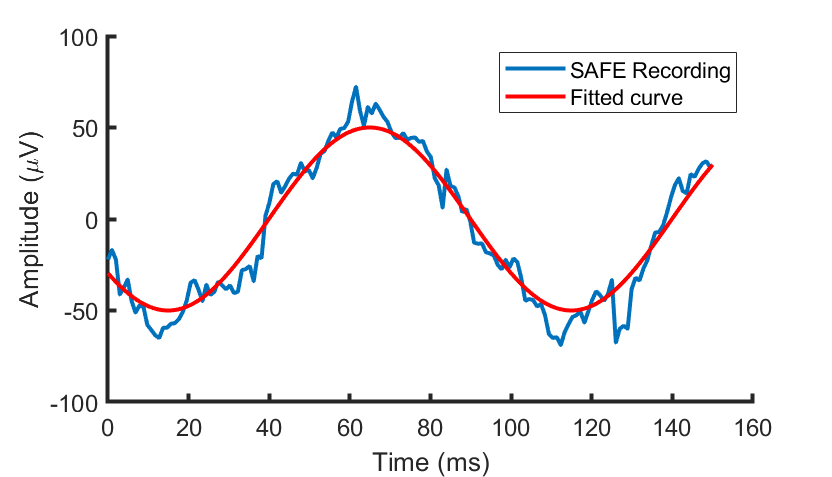}
    \caption[Fitted Sinusoid Example]{SAFE system recordings and a fitted sinusoid. The fitted curve is the signaled frequency and amplitude, with the algorithm only fitting for phase.}
    \label{fig:Fig2} 
\end{figure}

A high-pass filter with a 3~Hz cutoff frequency was applied to the data using the MATLAB highpass function, which uses a minimal order filter to achieve 60 dB attenuation in the stop band region. The data were segmented into epochs spanning one period. Some epochs contained artefact due to missed BLE packets. Epochs containing these artefacts were identified by eye and rejected. A sine wave with the delivered amplitude and frequency was fit to the data using the MATLAB fit function (Fig.~\ref{fig:Fig2}). The root-mean-squared error (RMSE) between the recorded wave and the generated signal was computed for both recording systems for comparison.

Since the recordings were  compared with a mathematical sine wave, it was important to consider the digital-to-analogue capabilities of the Analog Discovery as some noise in the comparison will be due to imperfections in the generated signal. The Analog Discovery has a step size of approximately 610~\si{\micro\volt\per\bit}. The step size seen by the amplifiers is reduced by the voltage divider to 3.39 \si{\pico\volt\per\bit}. The impact of this noise contribution is discussed in Section~\ref{Sec:CH2_SinResults}. We also report the packet loss by the SAFE system as the percentage of packets received of the total expected number of packets over a given recording session.

\begin{figure*}[b]
    \centering
    \begin{subfigure}[b]{0.35\textwidth}
        \includegraphics[width=\textwidth]{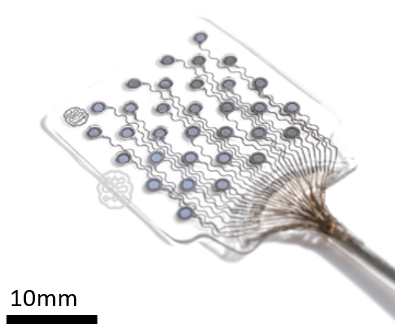}
        \caption{}
        \label{fig:Fig3_a}
    \end{subfigure}
    \hfill
    \begin{subfigure}[b]{0.6\textwidth}
        \includegraphics[width=\textwidth]{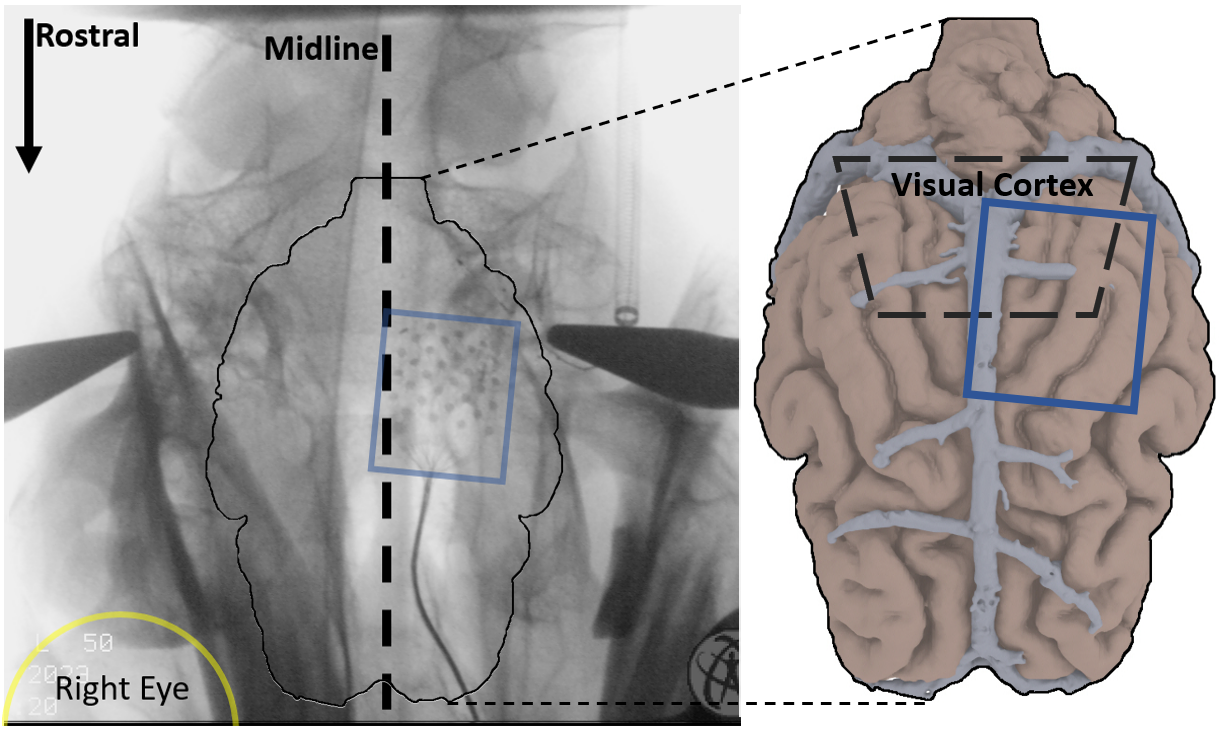}
        \caption{}
        \label{fig:Fig3_b}
    \end{subfigure}  
    \caption[VEP Experiment Setup]{(a) The CorTec AirRay Grid (image from: www.cortec-neuro.com, 2023) was used for sub-dural ECoG VEP recordings. (b) X-ray of subdural array placement in sheep (blue) was contralateral to stimulus of the eye (yellow). An approximate outline of the brain is overlayed on the xray and expanded to the right, depicting the location of the array over the visual cortex (image source: \cite{oxley_minimally_2016}).}
\end{figure*}

\subsubsection{Visual Evoked Potential (VEP) Recordings}
Both systems were also used to record visual evoked potentials (VEPs) with an electrocorticography (ECoG) array (CorTec, Germany) in four sheep. Fig.~\ref{fig:Fig3_a} shows the ECoG array. This experiment was approved by the Animal Ethics Committee at The Howard Florey Institute of Neuroscience and Mental Health (Approval number: 22010, 28 March 2022). 

The sheep were anaesthetised with isoflurane during the experiment. A craniotomy was performed and the subdural ECoG array was placed over the visual cortex in the left hemisphere (Fig.~\ref{fig:Fig3_b}). The VEPs were evoked using full field flash stimuli (Grass Instrument Co., USA) to the right eye of the animal. The stimulus was repeated at 0.99~Hz for 5 minutes for each amplifier. The g.USBAmp and SAFE device recordings both used six channels on the array proximal to the visual cortex, with the electrode on the array most distal to the visual cortex acting as the reference/ground. Recordings from the two devices were performed sequentially, first using the g.USBAmp system followed immediately by the SAFE system. Saline solution was applied to the eye prior to stimulation sessions to maintain eye health. 

The two recording systems interfaced with the trigger circuit in different ways, resulting in additional processing to ensure accurate comparison of the data. A photodiode connected to an Arduino (Nano 33 BLE, Arduino, Italy) acted as a trigger circuit. In the case of the g.USBAmp amplifier recording, the Arduino was connected to the amplifier trigger connector. When performing SAFE recordings, the Arduino used a serial connection to notify a PC (EliteBook x360 1030 G8 Notebook, HP, USA) when a flash event occurred. The PC ran a python script that simultaneously received BLE packets and serial trigger information, combined the two, and stored them in .csv format for offline analysis. In this way, the SAFE system could label the BLE packet during which the trigger occurred, but could not, unlike the g.USBAmp system, label the specific sample. This led to nonequivalent comparison of the VEP responses. To rectify this, we randomly (uniform distribution between 0-40~ms) shifted the g.USBAmp trigger times to simulate packets prior to analysis.

A comparison of the amplitude, signal-to-noise ratio (SNR), and VEP peak time between the two recording systems was used to evaluate their equivalence. Amplitude was calculated as the peak-to-peak voltage of the averaged-over-trials VEP signal between the time of the flash to 200~ms post-flash. SNR was computed as the ratio of the variance of the recorded signal segment from 0 to 200~ms post-stimulus, to the variance of the segment 0 to 200~ms pre-stimulus. The VEP peak time was manually chosen for each animal and channel as the time post-stimulus where peaks shared by both recording systems were observable when viewing the average-over-trials VEP signals. We present the mean and standard deviation of percent difference in VEP amplitude, SNR, and peak time across channels as an indicator of equivalence between the recording systems, where mean percent difference significantly different from zero indicates a difference in recording quality. 

\section{Results}
\label{sec:CH2_Results}

\subsection{Specifications}
The SAFE system was successfully fabricated to meet specifications. The full list of specifications, their design constraints, and the characteristics of the SAFE system with brief justifications are shown in Table~\ref{tab:CH2_Tab1}.

\begin{table*}[hbtp]
\centering
\footnotesize
\caption[SAFE System Specification List]{SAFE System Specification List. Constraints for each specification are designed to meet requirements of a chronic BCI implant. The specifications of the final SAFE system are listed, as are justifications for constraints and SAFE specifications.}
\label{tab:CH2_Tab1}
\begin{tabular}{@{}p{1cm} p{1.8cm} c c c p{9cm}@{}}
\toprule
\multicolumn{2}{c}{Specification}          & Constraint      & SAFE System       & Unit     & \multicolumn{1}{c}{Justification}                                                                                                                                                                                                                                 \\ \midrule
Sampling      & Rate                       & $\geq$500           & 1024              & Hz       & Maximum signal bandwidth is estimated as 120~Hz. 500~Hz is adequate to capture the signal, although 1024~Hz captures higher frequencies that help estimate the noise floor.                                                                                         \\ \cmidrule(l){2-6} 
              & Step size                  & $\leq$0.25          & 0.125             & µV/step  & \multirow{4}{9cm}{\justifying The signal amplitude is estimated to range from 10-100~µV. 12 bit resolution with 0.25~µV step size provides adequate signal resolution for amplitudes of this size, as well as large enough dynamic range to accommodate reasonable baseline noise fluctuations.} \\
              & & & & & \\ \cmidrule(l){2-5} 
              & Resolution                 & $\geq$12            & 16                & bits     &  \\
              & & & & & \\ \cmidrule(l){2-6} 
              & Inter-channel Sample Delay & $\leq$10            & 10                & µs       & Although simultaneous sampling is preferred, a short inter-channel sampling delay is tolerable.                                                                                                                                                                     \\ \midrule
Channels      & Number                     & 6$\leq$n$\leq$12      & 6                 & channels & 4 channels positioned over the motor cortex and 2 channels over the visual cortex allows for MI, SSVEP and P300 recording with a single implant.                                                                                                                  \\ \cmidrule(l){2-6} 
              & Location                   & Variable        & Variable          & -        & \multirow{2}{9cm}{\justifying Modular connection of electrode arrays allows for custom electrode types and locations to suit the user and their BCI needs and capabilities.} \\ \cmidrule(l){2-5}
              & Electrodes                 & Variable        & Variable          & -        &                                                                                                                                                                                                                                                          \\\midrule
Power         & Method                     & WPT             & Battery           & -        & Wireless power allows device operation without the need for cabling passing through the skin barrier, reducing risk of infection and discomfort. This method of power supply was not achieved in this project, as discussed later in Section~\ref{sec:Imp}.                \\ \cmidrule(l){2-6} 
              & Battery life               & \textgreater 16 & \textgreater 1000 & hours    & Battery life sufficient for continuous, daily use is required. The device draws 1.88 mA (during transmission) from a battery with 2600 mAh capacity.                                                                                                              \\ \midrule
Data Transfer & Medium                     & Wireless        & BLE               & -        & BLE is capable of transmitting data directly to the end device, rather than through a coil to an intermediate, externally worn device.                                                                                                                            \\ \cmidrule(l){2-6} 
              & Range                      & \textgreater 3  & 10                & m        & The device should be capable of controlling devices within the user’s vicinity.                                                                                                                                                                                   \\ \cmidrule(l){2-6} 
              & Transmit delay             & $\leq$50            & 40                & ms       & A short transmission delay provides fluid BCI control.                                                                                                                                                                                                            \\ \cmidrule(l){2-6} 
              & Packet loss                & $\leq$10            & 5±4               & \%       & 10\% packet loss of data sampled at 500~Hz is still providing adequate data for BCI control.                                                                                                                                                                      \\ \midrule
Form          & Implant size               & $\leq$20$\times$20$\times$5       & 12$\times$12$\times$3           & mm       & \multirow{2}{9cm}{\justifying The device (implant and external battery) should be small and light to minimise user's discomfort and aesthetic concerns.}                                                 \\\cmidrule(l){2-5} 
              & Weight                     & $\leq$100           & 63                & g        &                                                                                                                                                                                                                                                         \\ \bottomrule
\end{tabular}
\end{table*}

\begin{figure}[!t]
    \centering
    \begin{subfigure}[b]{0.48\textwidth}
        \includegraphics[width=\textwidth]{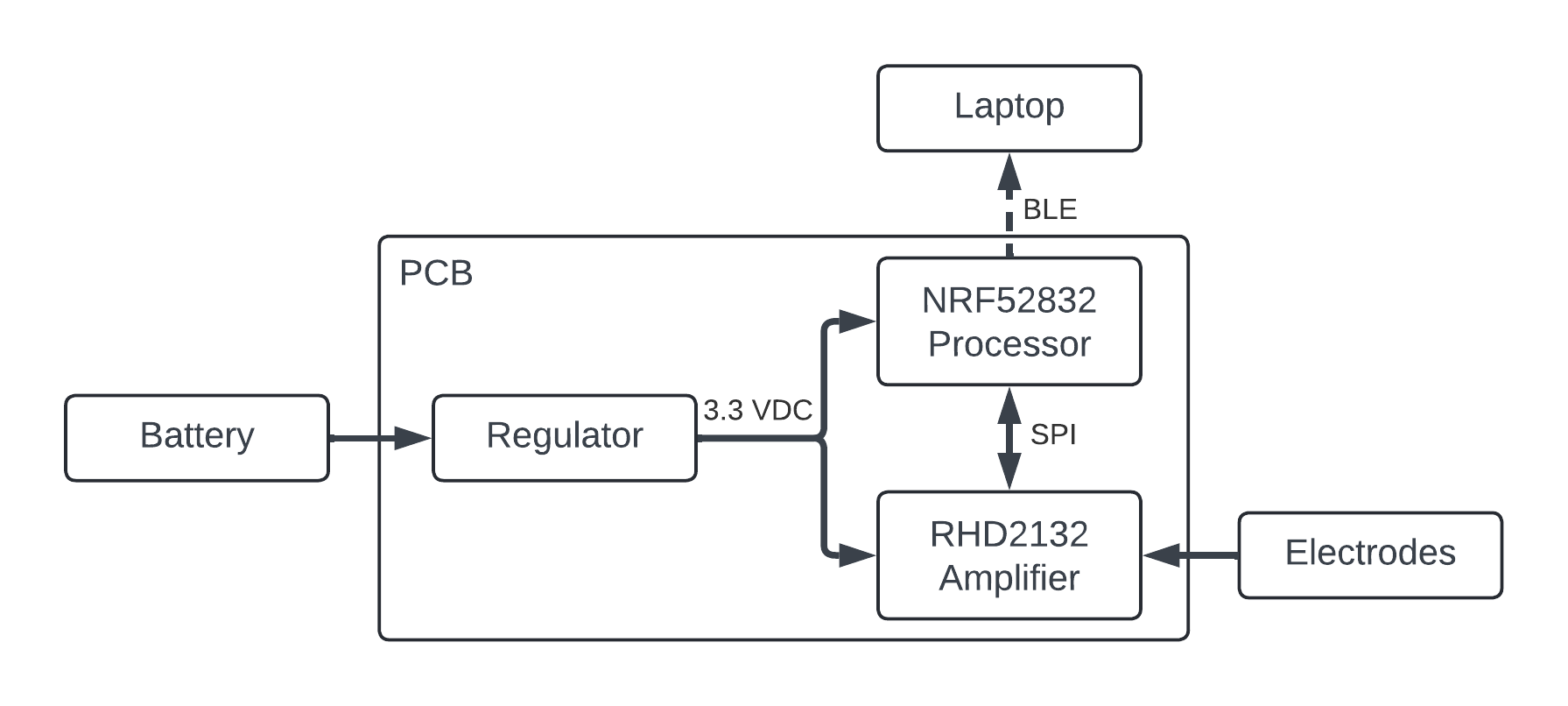}
        \caption{}
        \label{fig:Fig4_a}
    \end{subfigure}

    \begin{subfigure}[b]{0.48\textwidth}
        \includegraphics[width=\textwidth]{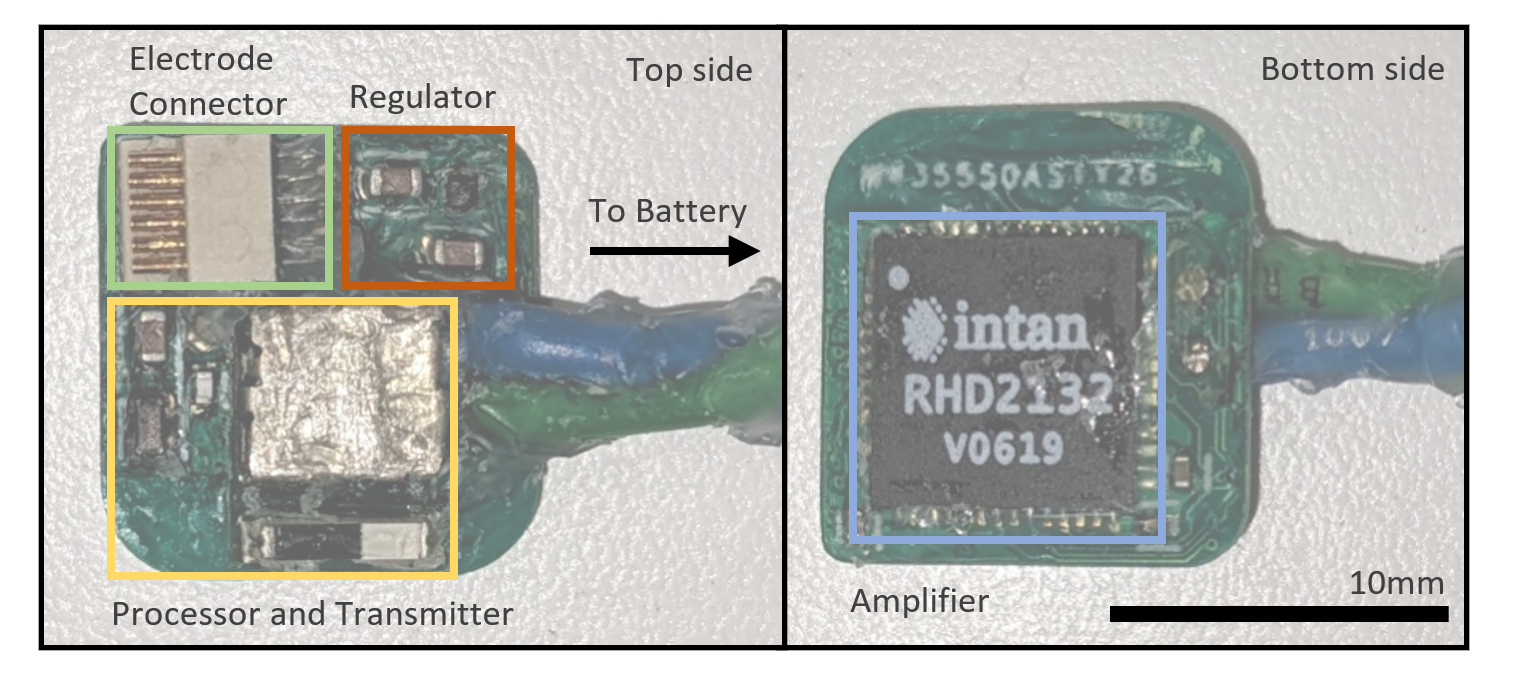}
        \caption{}
        \label{fig:Fig4_b}
    \end{subfigure}  
    \caption[SAFE System Schematic]{(a) The SAFE system schematiccomprising the custom PCB, battery, electrodes, and laptop computer for data receiving. (b) The two-sided PCB houses the power regulation, amplification and BLE transmission components of the device.}
\end{figure}

\subsubsection{System Overview}
A schematic of the SAFE system is shown in Fig.~\ref{fig:Fig4_a}. The system consists of the following components.
\begin{itemize}
    \item A 3.7~V lithium polymer 18650 battery cell (Core Electronics, Australia) with a 2600~mAh capacity was used to provide the device with ample supply for long-term recording sessions.  
    
	\item A custom printed circuit board (PCB) manufactured by PCBWay (China). Shown in Fig.~\ref{fig:Fig4_b}, the PCB comprises:
	
    \begin{itemize}
        \item a 3.3~V~DC linear regulator (NCP167AMX180TBG, Semiconductor Components Industries, LLC, USA). This model has a low dropout voltage and very small package size (4XDFN), occupying only 1~mm\textsuperscript{2}.
        
    	\item a processor module (MDBT42V-512KV2, Raytac Corporation, Taiwan) that includes a Nordic NRF52832 chip (Nordic Semiconductor, Norway) with BLE capabilities and chip antenna. The module has a small footprint (6.4×8.4~mm), BLE 5 with configurable throughput, low power modes, and an SPI interface for communication with the amplifier.
     
    	\item a multichannel electrophysiology amplifier chip (RHD2132, Intan Technologies, USA). This 9×9~mm amplifier includes 32 channels, potentially allowing for systems with much larger electrode arrays in future iterations.
     
    	\item an IO connector (Polarized Nano, Omnetics Connector Corporation, USA) for programming and six-channel electrode input. The electrode connector allows for the use of custom electrode arrays. The ground and reference channels are shorted.
     \end{itemize}
	
     \item A PC (EliteBook x360 1030 G8 Notebook, HP Inc., USA) for receiving and storing data from the implant via BLE with a python script implementing the Bleak package.
\end{itemize}

The system had the following specifications that were not met at this stage of development: trans-dermal data transmission and simultaneous sampling. Future iterations of the system will aim to incorporate these features.  

\subsection{Sinusoidal Recordings}
\label{Sec:CH2_SinResults}
\begin{figure*}[htbp]
    \centering
    \begin{subfigure}[b]{0.6\textwidth}
        \includegraphics[width=\textwidth]{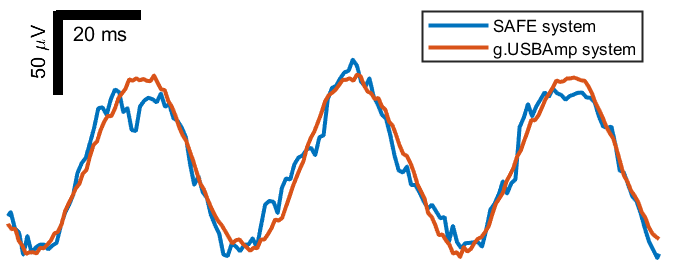}
        \caption{}
        \label{fig:Fig5_a}
    \end{subfigure}
    \hfill
    \begin{subfigure}[b]{0.39\textwidth}
        \includegraphics[width=\textwidth]{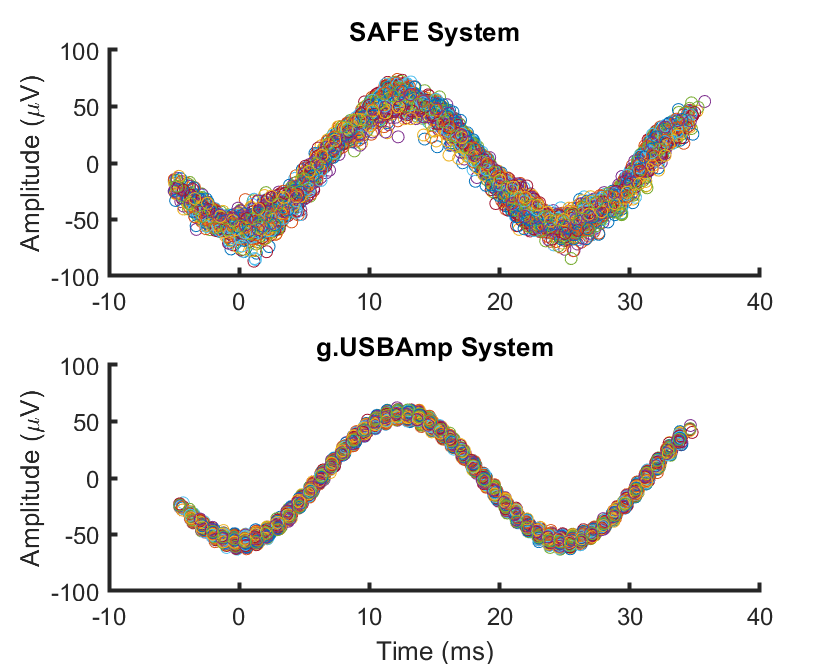}
        \caption{}
        \label{fig:Fig5_b}
    \end{subfigure}  

    \begin{subfigure}[b]{0.49\textwidth}
        \includegraphics[trim={2cm 0 2cm 0},clip, width=\textwidth]{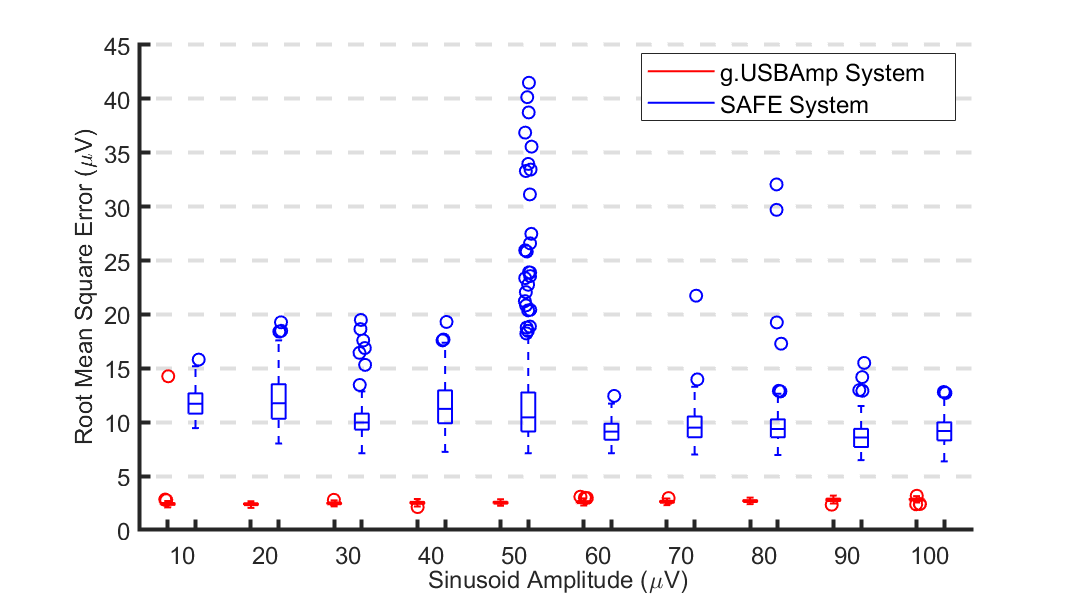}
        \caption{}
        \label{fig:Fig5_c}
    \end{subfigure} 
    \hfill\begin{subfigure}[b]{0.49\textwidth}
        \includegraphics[trim={2cm 0 2cm 0},clip, width=\textwidth]{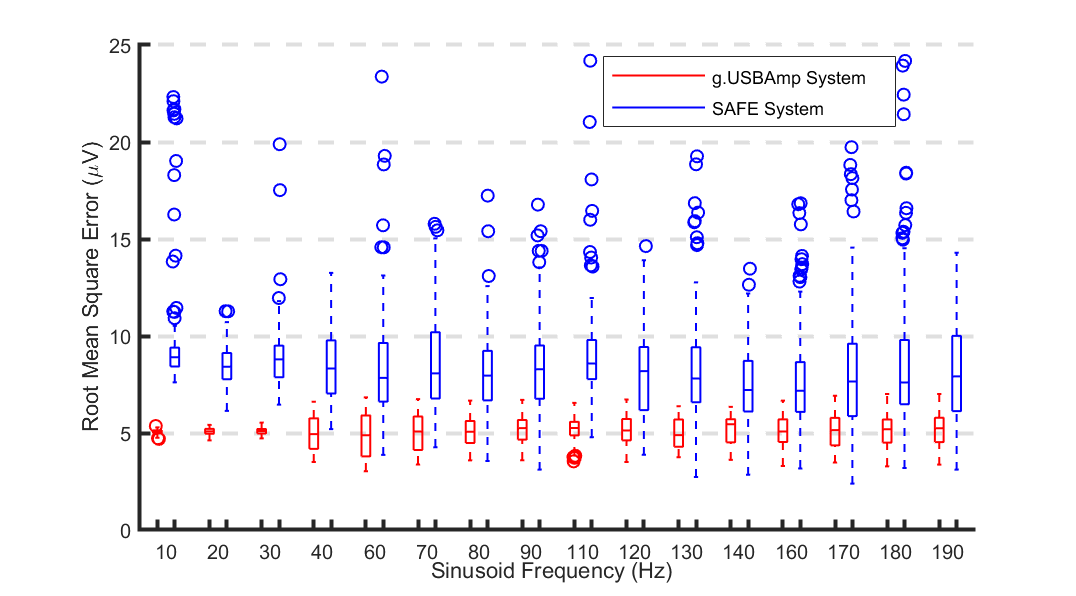}
        \caption{}
        \label{fig:Fig5_d}
    \end{subfigure}  
    \caption[Sinusoidal Recording Results]{(a) A comparison of raw traces from the SAFE (blue) and g.USBAmp (red) systems during generation of a 50~µV, 20~Hz sinusoid signal. (b) Aligned samples from the two amplifiers during 50~µV sinusoids with frequencies of 40~Hz. Samples by the SAFE system appear more spread than g.USBAmp recordings. (c)-(d) The RMSE from sinusoidal recordings using the SAFE and g.USBAmp systems. Each sample is one cycle. Each box consists of 200 samples. (c) All sinusoids had an amplitude of 50~µV with varying frequency. Mains line frequency (50~Hz) and harmonics were skipped. (d) All sinusoids had a frequency of 20~Hz with varying amplitude.}
\end{figure*}

A sample of raw data shown in Fig.~\ref{fig:Fig5_a} indicates that, while the SAFE system was able to provide an accurate estimation of the sinusoid, it was noisier than the g.USBAmp amplifier. Fig.~\ref{fig:Fig5_b} shows plots of every sample of the 40 and 90~Hz recordings, respectively, after been aligned in phase and overlayed, for both the SAFE (top row) and g.USBAmp (bottom row) systems. While the samples recorded using the SAFE system do not resemble as pure a sine wave as those of the g.USBAmp, the sine wave shape is still clearly defined.

The magnitude of the noise was estimated as the RMSE between the recordings and a sinusoid of the delivered amplitude and frequency, and a fitted phase. Fig.~\ref{fig:Fig5_c} and Fig.~\ref{fig:Fig5_d} show RMSE for the generated signals of different amplitudes and frequencies, respectively. There was a clear increase in noise in recordings from the SAFE system over g.USBAmp recordings. The error across all recordings was 9.4±3.3 and 4.0±1.4~µV for the SAFE and g.USBAmp systems, respectively (mean±std). The component of noise in the recorded signal due to quantisation by the Analog Discovery when generating the wave (3.39~\si{\pico\volt\per\bit}) was on the order of one thousandth of the total noise recorded, and as such was considered to have minimal impact. SAFE system packet loss was infrequent. Across the recordings of both frequency and amplitude varied sinusoidal waves, 95±4\% of the expected number of packets were received (mean±std).

\subsection{Visual Evoked Potential Recordings}
\begin{figure*}[htbp]
    \centering
    \begin{subfigure}[b]{0.24\textwidth}
        \includegraphics[width=\textwidth, trim={0 0.7cm 0 0},clip]{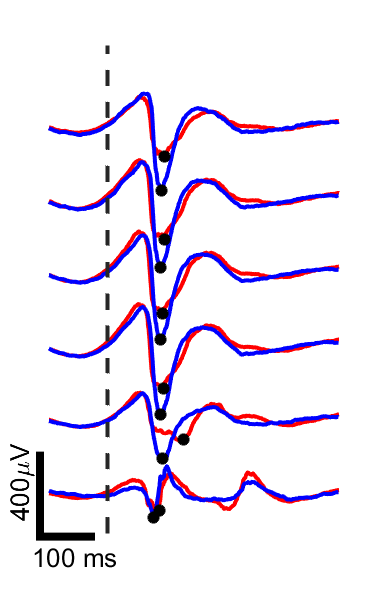}
        \caption{Sheep 1}
        \label{fig:Fig6_a}
    \end{subfigure}  
    \hfill
    \begin{subfigure}[b]{0.24\textwidth}
        \includegraphics[width=\textwidth, trim={0 0.7cm 0 0},clip]{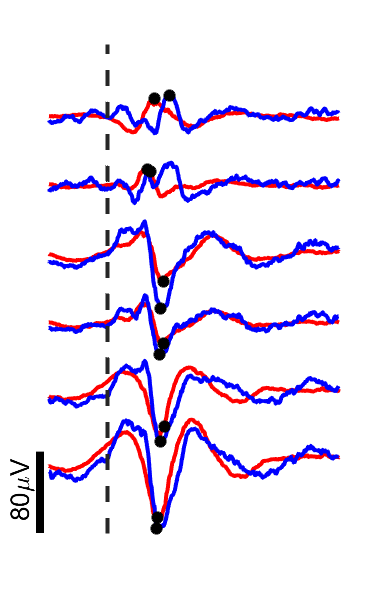}
        \caption{Sheep 2}
        \label{fig:Fig6_b}
    \end{subfigure} 
    \hfill
    \begin{subfigure}[b]{0.24\textwidth}
        \includegraphics[width=\textwidth, trim={0 0.7cm 0 0},clip]{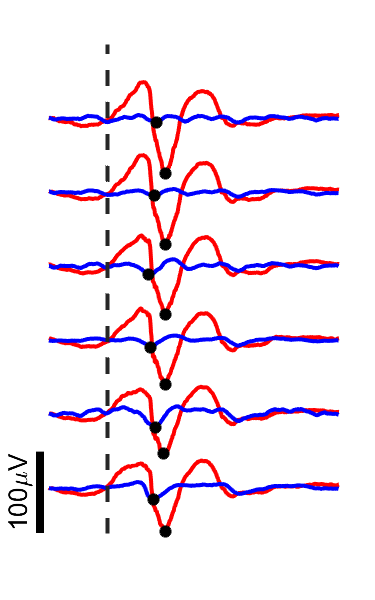}
        \caption{Sheep 3}
        \label{fig:Fig6_c}
    \end{subfigure}  
    \hfill
    \begin{subfigure}[b]{0.24\textwidth}
        \includegraphics[width=\textwidth, trim={0 0.7cm 0 0},clip]{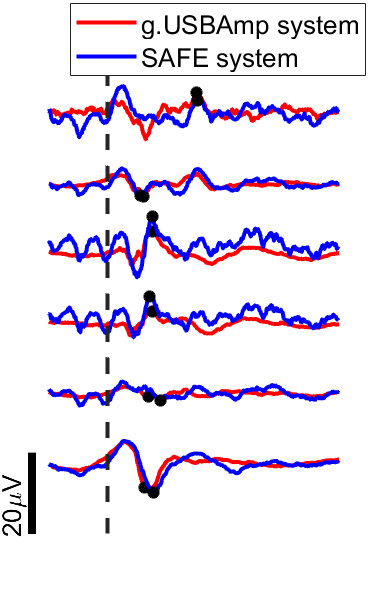}
        \caption{Sheep 4}
        \label{fig:Fig6_d}
    \end{subfigure} 

    \begin{subfigure}[b]{\textwidth}
        \includegraphics[width=\textwidth]{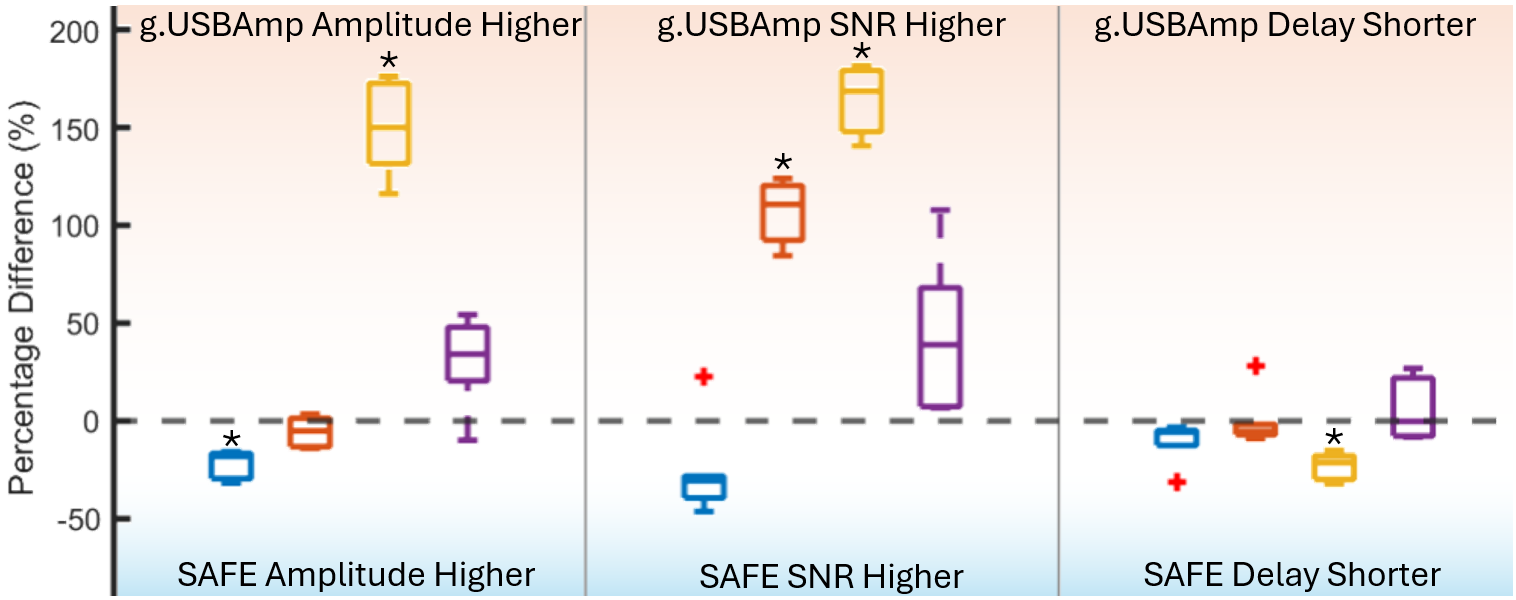}
        \caption{}
        \label{fig:Fig6_e}
    \end{subfigure} 
    \caption[VEP Responses Averaged-Over-Trials]{VEP response averaged over trials for each animal ((a)-(d)) from each channel. Responses recorded using the SAFE system are shown in blue and using the g.USBAmp amplifier in red. The dashed line indicates the time the flash occurred. Black dots indicate the manually chosen VEP peak time. (e) Percent difference between VEP amplitude, SNR, and peak time of the SAFE and g.USBAmp systems. Colours indicate different animals. Asterisks indicate where distributions are significantly different to zero (Student's t-test p$<$0.001).}
    \label{fig:Fig6}
\end{figure*}

Fig.~\ref{fig:Fig6} (a)-(d) depicts the ECoG VEP recordings in each sheep, averaged over trials. VEP amplitude, SNR, peak time, and the percent difference between both systems is presented in Table~\ref{tab:CH2_Tab2}. 25\% of the experiments recorded significantly greater VEP amplitude with the g.USBAmp system compared to the SAFE system  (p$<$0.001, Student's t-test). The SAFE system recorded sheep VEP responses with mean SNR greater than zero in all cases. 50\% of the experiments recorded significantly higher SNR with the g.USBAmp system than the SAFE system (p$<$0.001, Student's t-test). 75\% of the experiments showed no significant difference in the VEP peak time between systems (p$>$0.05, Student's t-test). Percent differences are illustrated in Fig.~\ref{fig:Fig6_e}.  

\begin{table}[htbp]
\centering
\footnotesize
\caption[VEP Response Comparison]{Comparison of VEP amplitude, SNR, and peak time (post stimulus), for recording device and sheep. Values are of the form mean$\pm$std across channels. Negative percent difference (d) indicates higher amplitude or SNR, or faster VEP peak recording by the SAFE system, whereas positive values indicate such outcomes for the g.USBAmp system. We include the p-values for Student's t-tests testing significance in the difference between percent difference values and zero. Bold percent differences are significantly different to zero (Student's t-test, p$<$0.001).}
\label{tab:CH2_Tab2}
\begin{tabular}{llcccc}
\hline
\multicolumn{1}{c}{}           &   & Sheep 1              & Sheep 2              & Sheep 3               & Sheep 4              \\ \hline
\multirow{4}{1.2cm}{Amplitude ($\mu$V)}     & SAFE    & 437±127        & 72±30          & 16±7            & 12±3           \\
                               & g.USBAmp   & 352±110        & 69±32          & 112±12          & 16±5           \\ \cmidrule{2-6}
                               & d & \textbf{-22±7} & -6±7           & \textbf{149±23} & 30±24 \\ 
                               & p-value & $<$0.001 & 0.12           & $<$0.001 & 0.029 \\ \hline
\multirow{4}{1.2cm}{SNR ($\mu$V)}           & SAFE    & 31±3           & 5±1            & 2±1             & 4±2            \\
                               & g.USBAmp   & 25±9           & 15±2           & 25±3            & 7±4            \\ \cmidrule{2-6}
                               & d & -26±25         & \textbf{107±16} & \textbf{164±17} & 44±39 \\
                               & p-value & 0.0511 & $<$0.001           & $<$0.001 & 0.037 \\\hline
\multirow{4}{1.2cm}{VEP Peak (ms)} & SAFE    & 91±6           & 90±13          & 80±5            & 90±34           \\
                               & g.USBAmp   & 102±15         & 90±10          & 100±2           & 85±35         \\ \cmidrule{2-6}
                               & d & -11±11         & 0±14           & \textbf{-23±7}  & 5±15           \\ 
                               & p-value & 0.0565 & 0.9597           & $<$0.001 & 0.4447 \\\hline
\end{tabular}
\end{table}

\section{Discussion}
Sub-scalp EEG addresses limitations of current BCI signal acquisition technologies by being less invasive than intracranial electrodes, not requiring daily donning and doffing procedure like scalp EEG, and being spatially versatile and removable unlike endovascular electrodes. However, existing sub-scalp EEG devices are designed primarily for seizure monitoring \cite{duun-henriksen_new_2020}, with a limited number of channels and sampling frequency that are not ideal for BCI applications. Here, we defined a list of specifications for a sub-scalp EEG devices suitable for BCI applications. The specification list encompassed requirements for data sampling and transmission, channel count, power consumption, size, and weight. With the exception of wireless power transmission, we successfully designed and fabricated a device to meet these specifications, the SAFE system, and validated its performance by way of comparison with a gold standard electrophysiology amplifier unit. Most notably, the device can transmit six channels of EEG data at 1024~Hz (per channel) wirelessly using BLE with minimal power consumption while being an appropriate size and weight for chronic sub-scalp implantation.

Comparison with a gold standard wired amplifier system (g.USBAmp) demonstrated that the commercial wired system offers higher SNR during sampling of generated sinusoids and VEPs in sheep models. However, as the SAFE system had a mean error of only 10~µV, the device is suitable for EEG recording considering EEG is typically in the order of 10-100~µV. Thus, while the SAFE system may not produce signal quality to the standard of the g.USBAmp, it is fit for purpose. This was evident in the VEP experiment, where the SAFE system showed a mean SNR greater than 0 dB in all cases, and no significant delay in response compared with the g.USBAmp system.

There are several potential reasons for the increased noise in the SAFE system compared with the g.USBAmp system; however, the exact cause is unclear. The SAFE PCB and cabling lacked the shielding of the g.USBAmp system. Additionally, the SAFE PCB design was not optimised to reduce coupling between digital and analogue signals, rather to minimise size for feasible implantation. Another drawback of such a small form factor is the limited ground plane. Each of these factors may have contributed considerably to the noise component recorded by the SAFE system. These factors are inherent limitations of implantable systems. Future investigations into sub-scalp EEG signal quality and suitability for BCI applications will use the SAFE device to ensure the data are of the quality we may expect from an implanted device, rather than laboratory equipment.

\subsection{Future Improvements}
Having demonstrated basic functionality, future iterations of the SAFE device should address current limitations of the system, particularly regarding data transmission, implantability, simultaneous sampling, and the receiver application.

\subsubsection{Implantable}
\label{sec:Imp}
Future iterations should seek to make the device implantable by including wireless power transmission, biocompatible encapsulation, and sterilisation methods, which are features that could not to be properly implemented in this project. The small form factor and low power consumption of the device support its feasibility as part of a chronically implanted BCI system. The PCB designed for this study could fit inside hermetically sealed containers currently used as part of cochlear implant systems \cite{uwe_baumann_device_2020, patrick_development_2006}, for example.

\subsubsection{Non-Simultaneous Sampling}
The amplifier chip does not sample all channels simultaneously. This results in sample skew and has several implications. Noise may not be consistent across all channels at the times of sampling. Some noise removal techniques estimate noise based on the common signal across channels (e.g., common average referencing) but, if the noise is not sampled simultaneously on each channel, removal in this way is less effective \cite{mcfarland_spatial_1997}. Additionally, some analysis techniques may consider signal phase (and variations in phase across spatial regions) to be an important feature \cite{wolpaw_braincomputer_2012, brunner_phase_2005}. Non-simultaneous sampling introduces a phase offset between samples, skewing this type of analysis. For these reasons, it is important to minimise the inter-channel sampling delay to reduce impact during data processing. 
    
The inter-channel sample delay for the SAFE system is approximately 10~µs. Voltage fluctuations would need to occur at frequencies in the order of 10-100~kHz or above to enact a significant change in voltage to the channels between samples. We do not expect to see significant noise at such high frequencies. Additionally, we expect the highest possible EEG signal bandwidth measurable with sub-scalp EEG to be less than 200~Hz. At this maximum bandwidth, a 10~µs inter-channel sampling delay constitutes a phase shift of just 1\% between sampling of the first and sixth channel. For these reasons, the effect of non-simultaneous sampling performed by the SAFE system is minimal. Despite this, simultaneous sampling components should be considered in future iterations to mitigate this limitation.

\subsubsection{Increased Channel Count}
Six-channel system can detect several BCI paradigms \cite{autthasan_single-channel_2020, cecotti_robust_2011, colwell_channel_2014, juan_alberto_brain-computer_2021, karunasena_single-channel_2021, mccann_electrode_2015, speier_method_2015, xu_channel_2013, lun_simplified_2020}. These paradigms have demonstrated high functioning BCI systems that can control a cursor in 2D \cite{kayagil_binary_2009, wolpaw_control_2004} and 3D  \cite{mcfarland_electroencephalographic_2010} space, type \cite{chen_high-speed_2015}, operate wheelchairs \cite{al-qaysi_review_2018}, or control robotic prosthesis \cite{karunasena_single-channel_2021}. However, more channels may be useful for chronic systems. EEG patterns have been shown to vary within subjects both over short times due to fatigue \cite{qi_neural_2019}, attention \cite{ray_high-frequency_2008}, stress \cite{perez-valero_quantitative_2021}, emotion \cite{yang_high_2020}, hunger \cite{hoffman_eeg_1998}, and pain \cite{zis_eeg_2022}. A BCI would ideally function irrespective of these factors. Adaptive BCIs that can respond to changes in the user’s state or the environment may become increasingly important as BCI moves away from acute lab experiments and into the homes of users. These systems may benefit greatly from higher channel counts, providing spatial information that can help isolate underlying features of interest in the presence of these artefacts, thereby increasing the robustness of the BCI and user satisfaction \cite{lee_individual_2022, schreiner_mapping_2024}. However, an increase in channels may demand higher power and transmission throughput, and complicates the sub-scalp implantation procedure. Future work should endeavor to determine the optimal system that balances these features for long-term in-human BCIs.

\subsubsection{Receiver Data Handling}
On the receiver end, additional functions that allow for modular, online processing and interfacing of the EEG data with various applications, as well as implementing systems for the secure handling of the user's biological information, would elevate this device to an entire BCI system. End effectors, including cursors, keyboards, phones, wheelchairs, prostheses, gaming consoles, televisions, and other devices, including those that are part of the internet of things, should be considered based on the interests of the user and their support network. Developing the system as a platform technology, whereby users and their support networks are able to develop their own applications, would see the device used in ways beyond originally envisioned. In this way, the technology will provide the most impact across of cohort of users with a broad variety of conditions, needs, and interests.

\subsection{Considerations for the Future of Sub-Scalp BCI}
The advent of continuous EEG streaming from a potentially large cohort of users provides the opportunity for several interesting research avenues, both within and outside the context of BCI. Large banks of sub-scalp BCI data may permit the training of sophisticated AI algorithms, allowing these minimally invasive systems to approach levels of functional performance previously only thought possible with ECoG or penetrating electrodes. Data gathered across large cohorts of users could be used to pre-train BCI classifiers, reducing the amount of data required to gather from new users to calibrate their own personal classifiers, thereby reducing training time for users from months or weeks to days or even hours. 

In addition to BCI, high throughput sub-scalp EEG could also monitor other conditions commonly experienced by target users, such as pain \cite{zis_eeg_2022}, fatigue \cite{qi_neural_2019}, stress \cite{perez-valero_quantitative_2021}, depression \cite{neto_depression_2019}, apnea \cite{zhao_classification_2021}, and so on. Providing this information to clinicians may result in better health outcomes for users. In some cases, monitoring these conditions may provide enough benefit to support implantation of sub-scalp EEG devices in persons without severe paralysis, as in the case of long-term seizure monitoring for persons with epilepsy \cite{stirling_seizure_2021}. 

Perhaps most importantly, the preferences of users (with various conditions), carers, and clinicians regarding sub-scalp BCI should be investigated. Previous studies have introduced such groups to the cumbersome non-invasive EEG systems and invasive intracranial arrays, surveying their views regarding BCI performance, aesthetics, effort, and other such parameters \cite{kubler_user-centered_2014, miralles_brain_2015, rashid_current_2020, miralles_braincomputer_2015, mcfarland_eeg-based_2017}. However, whether sub-scalp BCI would appeal to such groups as an attractive alternative is unclear. These stakeholders should provide continuous input during sub-scalp BCI development to ensure the resulting product is fit for purpose and provides maximum benefit.

Prior to moving to human studies, more questions must be answered regarding the feasibility of sub-scalp BCI. Having developed a miniature device that is capable of sampling and transmitting EEG data, future studies will use the device to examine sub-scalp EEG quality in animal models, particularly in recording neural activity useful for BCI applications. Key insights to consider before moving to human models include understanding the SNR, spatial resolution, and bandwidth of sub-scalp EEG relative to other BCI signal acquisition methods, such as ECoG and EV electrodes, as well as the impact of positioning the sub-scalp electrodes at different depths in the sub-scalp space (e.g., above the periosteum or flush with the skull). 

\section{Conclusion}
Sub-scalp EEG addresses limitations of current BCI signal acquisition methods. However, current sub-scalp EEG devices are designed for chronic seizure monitoring, and are not suitable for BCI applications. Here, we have formulated a list of specifications for sub-scalp BCI devices that addresses requirements such as channel count, sampling frequency, power consumption, and form factor. We fabricated the SAFE device, an electrophysiology amplifier and wireless transmitter system designed to meet these specifications, and validated the system by way of comparison with a state of the art, commercially available amplifier. The system features six-channels, 1024~Hz sampling, BLE wireless data transmission, and a compact form factor of 12$\times$12~mm. The SAFE system demonstrated low noise recording of generated sinusoid signals and ECoG VEPs in sheep models. Going forward, the device will be used in animal experiments to assess signal quality of sub-scalp EEG for the purposes of BCI applications. Data collected from these experiments will build the foundation for future in-human sub-scalp BCI studies, propelling this technology toward chronic, in-home BCI applications that improve quality of life for those living with physical disability.

\section{Acknowledgments}
We thank research assistant Huakun Xin, veterinary technician Tomas Vale, and animal handler Quan Nguyen (The Florey Institute of Neuroscience and Mental Health) for their assistance in caring for the animals, experiment preparations, and monitoring during the animal experiments.

We also wish to thank the animals used for this research. Animal studies are crucial for the realisation of medical devices. We intend to ensure insights gained from their sacrifice lead to improved quality of life for persons with severe disability.

\bibliographystyle{IEEEtran} 
\bibliography{ieeeRefs.bib}

\end{document}